\documentstyle[prd,aps,multicol,eqsecnum,amssymb]{revtex}
\begin{document}
\draft 

\title{ An action for supergravity interacting 
with super-p-brane sources}

\author{Igor A. Bandos$^{\ast,\ddagger}$, Jos\'e A. de 
Azc\'arraga$^{\ast}$,
Jos\'e M. Izquierdo$^{\dagger}$ and Jerzy Lukierski$^{\ast ,\star}$} 
\address{$^{\ast}$Departamento de F\'{\i}sica Te\'orica and IFIC,
 46100-Burjassot (Valencia), Spain} 
\address{$^{\dagger}$Departamento de F\'{\i}sica Te\'orica,
 Facultad de Ciencias, 47011-Valladolid, Spain}
\address{$^{\ddagger}$Institute for Theoretical Physics, NSC KIPT, 
UA61108,
Kharkov, Ukraine} 
\address{ $^\star$Institute for Theoretical  Physics,
pl. Maxa Borna 9, 50-204 Wroclaw, Poland}
\date{FTUV/01-2404, IFIC/01-22, April 24, 2001}
\maketitle 

\def\theequation{\arabic{equation}}

\begin{abstract}
We describe the coupled system of supergravity and a superbrane source 
by the sum of the group manifold action 
for  D--dimensional supergravity 
and the action for a super--$p$--brane. 
We derive the generalized  
Einstein equation with the source  
and discuss the local fermionic symmetries of the coupled action. 
Our scheme could be especially relevant  in $D=11, 10$, in which 
the superfield actions for supergravity are not known.

\end{abstract}

\pacs{PACS numbers: 11.30.Pb, 11.25.-w, 04.65.+e, 11.10Kk}

\begin{multicols}{2}

\narrowtext

{\bf 1.} {\it Introduction}. 
The interaction of superbranes with supergravity is usually  studied 
either in the 
bosonic approximation 
(see {\it e.g.} \cite{SUGRA,bbs,ds}) or by means of worldvolume actions 
\cite{BST,AETW,c0,blnpst} 
written in a supergravity background. 
In the latter approach the requirement of $\kappa$--symmetry of 
the superbrane action leads to the superfield supergravity constraints. 
However, for $D=11,10$ these turn out to be  on--shell constraints  
whose consistency implies  
{\sl free} supergravity equations without any source. 
Thus, the existing approaches 
deal only with an approximate description of the supergravity---superbrane 
coupled system. 

The natural supersymmetric extension 
of the interaction between the bosonic 
fields of the supergravity supermultiplet and bosonic branes 
is provided by the sum 
of the superbrane action and the full action for supergravity. 
The main problem for this extension is  
that the superbrane actions involve the 
supergravity background in the {\sl  superfield} formulation, while 
the actions for 
$D=11,10$ supergravity are known in  
{\sl component} form only. Thus, the component action 
for supergravity and the supergravity background 
in the superbrane action are written in terms of different variables 
\cite{11SG,ds}.

The first message of this letter is that a supergravity formulation 
allowing for a description of 
supergravity coupled to a dynamical superbrane source 
may be provided by the {\sl group manifold} (GM) {\sl action} 
\cite{rheo}.
The key point is that this action is formulated in terms of superfields, 
but in which the Grassmann 
coordinate ${\theta}^{ {\alpha}}$ of curved superspace  is replaced by 
the fermionic 
field $\tilde{\theta}^{ {\alpha}}(x)$.  
The same superfields also enter in the super--$p$--brane action, where now 
the bosonic and fermionic coordinates $x^\mu$, $\theta^\alpha$, are replaced 
by worldvolume fields $\hat{x}^\mu (\xi)$,  
$\hat{\theta}^{\alpha}(\xi)$.  Thus, both the GM supergravity and superbrane 
actions depend on the same `superfield' variables and 
the joint variational problem is well posed. 

We shall derive in this framework  
the supergeneralization of the Einstein equation with a source and show that  
on shell the superbrane field $\hat{\theta}^{ {\alpha}}(\xi)$ 
coincides 
with the restriction  $\tilde{\theta}^{ {\alpha}}(\hat{x})\equiv 
\tilde{\theta}^{ {\alpha}}(\hat{x}(\xi)) $
of the supergravity fermionic field $\tilde{\theta}^{ {\alpha}}(x)$.
We shall also describe the local fermionic symmetries of the coupled 
system.   

\medskip

{\bf 2.} {\it Action for the coupled system  
of supergravity and a super--$p$--brane source}. 
Let $Z^M$=$(x^{ {\mu}}, {\theta}^{ {\alpha}})$ be the coordinates of 
superspace ${\Sigma}^{(D|n)}$ 
(${\mu}=0,1,...,(D-1)$,  
 ${\alpha}=1,..., n$, $n=2^{[D/2]}N$)
and let $\tilde{\phi}$ 
[$\hat{\phi}$] map
spacetime $M^D$ [worldvolume $W^{p+1}$ parametrized by 
$\xi^m=(\tau , \sigma^1, \ldots \sigma^p)$, 
$m=0,\ldots, p$] into ${\Sigma}^{(D|n)}$. Then,  
\begin{eqnarray}\label{MD}
& \tilde{\phi}: M^{D}\rightarrow {\Sigma}^{(D|n)}\; , 
\quad x^\mu \mapsto  \tilde{Z}^{ {M}}(x)=   
(x^{ {\mu}}, \tilde{\theta}^{ {\alpha}}(x))\; ; 
\\ 
\label{M1}
& 
\hat{\phi}: W^{p+1}\rightarrow {\Sigma}^{(D|n)}\; , 
\;\; \xi^m \mapsto \hat{Z}^{ {M}}(\xi )=   
(\hat{x}^{ {\mu}}(\xi), \hat{\theta}^{ {\alpha}}(\xi) ) \; . 
\end{eqnarray} 
Eqs. (\ref{MD}), (\ref{M1}) determine $D$-- and $(p+1)$--dimensional 
surfaces ${\cal M}^D$ and ${\cal W}^{p+1}$ in 
superspace  $\Sigma^{(D|n)}$.

The action is assumed to be of the form 
\begin{equation}\label{SSG+p}
S= \int_{M^{D}}  \tilde{{\cal L}}_{D}
+  \int_{W^{p+1}}  \hat{{\cal L}}_{p+1} \equiv S_{D,SG}+ S_{D,p} \; ,   
\qquad 
\end{equation}  
where 
$
\tilde{{\cal L}}_{D}
$ 
is the Lagrangian $D$--form (on $M^D$) of 
the GM action for $D$--dimensional supergravity 
(the explicit form of which is known for the $D=4,6$ and $D=11$ cases 
only \cite{rheo}),  
and 
$\hat{{\cal L}}_{p+1}$ is the Lagrangian ($p$+$1$)--form (on $W^{p+1}$) 
of the super-$p$-brane.

{\bf 2a.} The 
{\it Lagrangian form 
$\tilde{{\cal L}}_{D}= {\cal L}_{D}(\tilde{Z}, d\tilde{Z})$
of the} GM {\it action $ S_{D,SG}=\int_{M^D}\tilde{{\cal L}}_{D}$ 
for supergravity}
 \cite{rheo} is the 
pull--back $\tilde{{\cal L}}_{D}=\tilde{\phi}^*({\cal L}_{D})$ 
to ${M}^{D}$ of a $D$-form ${{\cal L}}_{D}$ on  ${\Sigma}^{(D|n)}$, 
\begin{equation}\label{LD}
{\cal L}_{D}  (Z;dZ)= {1\over D!} dZ^{ {M}_D}\ldots 
dZ^{ {M}_1} {\cal L}_{ {M}_1 \ldots  {M}_D}(Z) \; .
\end{equation} 
${\cal L}_{D}$ is constructed from the supervielbein 
\begin{equation}\label{EA}
E^{ {\cal A}}= (E^{ {a}}, E^{\underline{\alpha}}) = 
dZ^{ {M}} E_{ {M}}^{~ {\cal A}}(Z) \; ,  \qquad 
\end{equation} 
the torsions ${T}^{\underline{\alpha}}= $ 
${\cal D}{E}^{\underline{\alpha}} =$ 
$d{E}^{\underline{\alpha}}-  {E}^{\underline{\beta}} 
 {w}_{\underline{\beta}}^{~\underline{\alpha}}$, ${T}^{ {a}}=$  
 $d{E}^{ {a}}-  {E}^{ {b}}  
 {w}_{ {b}}^{~ {a}}$,    
the curvature $ R^{ {a} {b}}= dw^{ {a} {b}}- w^{ {a} {c}}
w_{ {c}}^{~ {b}}
$
of the (independent) spin connection  
$w^{ {a} {b}}= dZ^{ {M}} w_{ {M}}^{ {a} {b}}=-w^{ {b} {a}}$, 
${w}_{\underline{\beta}}^{~\underline{\alpha}}=1/4 w^{ {a} {b}} 
\Gamma_{ {a} {b}}{}_{\underline{\beta}}^{~\underline{\alpha}}$ and, possibly, 
from a set of superforms 
$C_{p+1} \equiv {1/(p+1)!} dZ^{ {M}_{p+1}} \ldots 
dZ^{ {M}_1} C_{ {M}_1 \ldots  {M}_{p+1}}(Z)$ 
providing superfield extensions 
of the antisymmetric gauge fields ${C}_{ {\mu}_1 \ldots  {\mu}_{p+1}}(x)$
entering the supergravity multiplet ($C_3$ for $D=11$ supergravity), 
supplemented by the auxiliary tensor zero--form 
fields  $F^{a_1\ldots a_{p+2}}$ 
($F^{ {a} {b} {c} {d}}$ for $D$=$11$ supergravity) that are 
necessary to rewrite the kinetic terms for the gauge field 
{\sl without using} the Hodge $*$-operator. 
The most important  
terms in ${\cal L}_D$ are 
\begin{eqnarray}\label{LD1}
& {\cal L}_D = R^{ {a} {b}} E^{\wedge (D-2)}_{ {a} {b}} 
-{2i\over 3} {\cal D} E^{\underline{\alpha}} ~E^{\underline{\beta}}~ 
E^{\wedge (D-3)}_{ {a} {b} {c}}~
{\Gamma}^{ {a} {b} {c}}_{\underline{\alpha}\underline{\beta}} 
\nonumber \\ & +\ldots \quad , 
\end{eqnarray} 
where 
$ E^{\wedge (D-q)}_{ {a}_1\ldots  {a}_q}  \equiv 
{1 \over (D-q)!} {\varepsilon}_{ {a}_1\ldots  {a}_q {b}_1\ldots  {b}_{D-q}}
$ $ E^{ {b}_1} \ldots  E^{ {b}_{D-q}} $.

Since 
$\tilde{\phi}^*(d\theta^{\alpha})=d\tilde{\theta}^{ {\alpha}}(x) =
dx^{ {\mu}} \partial_{ {\mu}} \tilde{\theta}^{ {\alpha}}(x)$,  
\begin{equation}\label{tLD}
 \tilde{\cal L}_{D}= {\cal L}_{D} (x,\tilde{\theta};dx,d\tilde{\theta})
= d^D x \, L_{SG} (x,\tilde{\theta},\partial_{ {\mu}}\tilde{\theta}) \; , 
\end{equation}
where $d^Dx$ is the volume $D$--form on $M^D$. 
Hence, the GM action for {\sl free} supergravity 
can be treated  as a usual (component) action, but with a  
dependence on  $\tilde{\theta}(x)$
determined by the map $\tilde{\phi}^*$ (eq. (\ref{MD})).  
Such a dependence on $\tilde{\theta}(x)$ allows one to treat all the 
equations of 
motion, obtained by varying (\ref{tLD}), as {\sl superfield equations valid in 
superspace ${\Sigma}^{(D|n)}$}. 
This is because the superdiffeomorphism invariance of 
${\cal L}_{D}$ implies that 
 $\tilde{\cal L}_{D}$ is invariant under the 
pull-back of superdiffeomorphisms 
\begin{eqnarray}\label{susy}
\delta_{\tilde{s}}\tilde{Z}^M = b^M (x) \, : \;   
\delta_{\tilde{s}} x^\mu= b^\mu (x) \, , \; 
\delta_{\tilde{s}} \tilde{\theta}^{\alpha}(x)= {\varepsilon}^{\alpha}(x)
\, ,
\end{eqnarray}
plus the corresponding `superfield' transformations 
$\delta_{\tilde{s}}^\prime {E}_M^A(\tilde{Z})$ := 
${E}_M^{\prime A}(\tilde{Z}(x))
- \tilde{E}^{A}_M(\tilde{Z}(x))$, 
$\delta_{\tilde{s}}^\prime w_M^{ab}(\tilde Z)$,
etc. As a result, $\delta S_{D,SG}/\delta  \tilde{\theta}$ 
does not produce an independent equation of motion and  
the field equations do not depend 
on the choice of the surface ${\cal M}^D$, {\it i.e.} of $\tilde{\phi}$. 
As the union of all such surfaces 
covers ${\Sigma}^{(D|n)}$, one has a reason to lift 
the field equations to superspace 
\cite{rheo1} by omitting the pull--back from the 
superforms.

The equations of motion resulting from  
$\delta \tilde{w}^{ab}$, $\delta \tilde{F}^{a_1 ... a_{p+2}}$, 
$\delta \tilde{C}_{p+1}$, 
$\delta \tilde{E}^{\underline{\alpha}}$ and 
$\delta \tilde{E}^{a}$ variations are 
\begin{eqnarray}
\label{tTa}
& \tilde{T}^a  + i \tilde{E}^{\underline{\alpha}} 
\tilde{E}^{\underline{\beta}} 
\Gamma^a_{\underline{\alpha}\underline{\beta}}=0 \; , \\ 
\label{tH}
& d\tilde{C}_{p+1} - c\,   \tilde{E}^{\underline{\alpha}} 
\tilde{E}^{\underline{\beta}} \tilde{E}^{a_1} \ldots 
\tilde{E}^{a_p} 
\Gamma_{a_1\ldots a_p} {}_{\underline{\alpha}\underline{\beta}} + \nonumber 
\\ & \quad   
+ {1\over (p+2)!} E^{a_{p+2}}\ldots E^{a_1}
\tilde{F}_{a_1\ldots a_{p+2}} = 0 \; , 
\\ 
\label{tG} 
&\tilde{G}_{(D-p-1)}:=
d (\tilde{E}^{\wedge (D-p-2)}_{a_1\ldots a_{p+2}} 
\tilde{F}^{a_1\ldots a_{p+2}})
+ \ldots=0 \; ,
\\ 
\label{tRS}
& {\tilde{\Psi}}_{(D-1)\underline{\alpha}} := {{4i\over 3}}{\cal D} 
\tilde{E}^{\underline{\beta}} \,  
\tilde{E}^{\wedge (D-3)}_{ {a} {b} {c}}\, 
{\Gamma}^{ {a} {b} {c}}_{\underline{\beta}\underline{\alpha}} 
+\ldots =0 \; , \\ 
  \label{tMD-1}
& \tilde{M}_{(D-1)~{ {a}}} := \tilde{R}^{bc} \, 
\tilde{E}^{\wedge (D-3)}_{abc} + 
\ldots =0 \; , 
\end{eqnarray} 
where $c$ is a constant depending on $D$ and $p$. 
After their lifting to superspace, eqs. (\ref{tTa}), (\ref{tH}) 
give rise to the {\sl free} supergravity constraints on 
$\Sigma^{(D|n)}$  (see \cite{rheo})
\begin{eqnarray}\label{Ta}
&T^a = -i E^{\underline{\alpha}} ~  E^{\underline{\beta}} 
\Gamma^a_{\underline{\alpha}\underline{\beta}}\; , \\ 
\label{H} 
& d{C}_{p+1} =  c\, {E}^{\underline{\alpha}} 
{E}^{\underline{\beta}} {E}^{a_1} \ldots 
{E}^{a_p} 
\Gamma_{a_1\ldots a_p} {}_{\underline{\alpha}\underline{\beta}}\; 
+ \ldots \; ,
\end{eqnarray}
while eqs. (\ref{tG}), (\ref{tRS}) and (\ref{tMD-1})  
become   
the supergeneralization of the free gauge field, 
Rarita--Schwinger and Einstein 
equations on $\Sigma^{(D|n)}$,  
\begin{eqnarray}
\label{G}
&{G}_{(D-p-1)}:= 
d ({E}^{\wedge (D-p-2)}_{a_1\ldots a_{p+2}} F^{a_1\ldots a_{p+2}})
+ \ldots=0 \; ,\\
\label{RS}
& {{\Psi}}_{(D-1)\underline{\alpha}}
:= {{4i\over 3}}{\cal D} 
{E}^{\underline{\beta}} \, 
{E}^{\wedge (D-3)}_{ {a} {b} {c}}~
{\Gamma}^{ {a} {b} {c}}_{\underline{\beta}\underline{\alpha}} 
+\ldots = 0\; , \\  
\label{MD-1}
& M_{(D-1)~{ {a}}}:= {R}^{bc} \, 
{E}^{\wedge (D-3)}_{abc} + 
\ldots =0 \; .  
\end{eqnarray}
\indent Moreover, 
the Lagrangian form ${\cal L}_D$ 
is chosen in such a way that $\int\tilde{\cal L}_D$ 
possesses the local supersymmetry  \cite{rheo}
\begin{eqnarray}\label{lsusy}
& \delta_{ls} \tilde{E}^a = -2i \tilde{E}^{\underline{\alpha}} 
\Gamma^a_{\underline{\alpha}\underline{\beta}}
\epsilon^{\underline{\beta}}(x,\tilde{\theta}(x))\; ,\quad  
\\   
& \delta_{ls}\tilde{E}^{\underline{\alpha}} = 
{\cal D} {\epsilon}^{\underline{\alpha}}(x,\tilde{\theta}(x))+ \ldots \; ,\quad
\delta_{ls} \tilde{w}^{ab}= \ldots \; , \quad  \ldots \; , 
 \nonumber \\ \label{lsusyx} & 
\delta_{ls} x^\mu=0 \; ,  \quad \delta_{ls}\tilde{\theta}^{\beta}(x)=0 \; , 
\end{eqnarray}
where the dots in the expression for 
$\delta_{ls} \tilde{E}^{\underline{\alpha}}$,  $\delta_{ls} \tilde{w}^{ab}$
denote the terms including ${\epsilon}^{\underline{\alpha}}$ 
without derivatives 
and ${\epsilon}^{\underline{\alpha}}(x,\tilde{\theta}(x))= 
\tilde{\phi}^*({\epsilon}^{\underline{\alpha}}(x,{\theta}))$. 
In the framework of the second Noether theorem,  
the symmetry (\ref{lsusy}) is reflected by the (Noether) identity 
\begin{equation}\label{DRS}
{\cal D} \tilde{\Psi}_{(D-1)\underline{\alpha}} - 
2i \tilde{M}_{(D-1) a}\, \tilde{E}^{\underline{\beta}}
\Gamma^a_{\underline{\alpha}\underline{\beta}} + \ldots \equiv 0\; ,  
\end{equation}
which, as well as ${\cal D} \tilde{M}_{(D-1) a}- \ldots \equiv 0$,  
 follow from the definitions for 
$\tilde{\Psi}$ and $\tilde{M}$ in 
(\ref{tRS}),  
(\ref{tMD-1}), see \cite{rheo}, and where   
the terms denoted by dots 
are proportional to the expressions (\ref{tTa})--(\ref{tRS}). 
In contrast with (\ref{susy}), 
this local supersymmetry 
survives after setting 
$\tilde{\theta}(x)$=$0$ because of (\ref{lsusyx}).

The lifting of the equations of motion to  superspace 
fails when  $\delta\tilde{\theta}$ produces an independent equation 
({\it cf.} \cite{VS73}). 
It may also fail when  $\delta S/\delta\tilde{\theta}\equiv 0$   
if $\tilde{\theta}$ appears explicitly in an independent equation of 
motion or when the lifting of some equation of motion to 
 superspace results in a contradiction (as  
it will be the case in Sec. 3 
for the interacting system). 
This happens as well in the model with self--dual gauge fields 
({\it cf.}  \cite{bpst}).

{\bf 2b.} The {\it Lagrangian form for a  super--p--brane}  
is \cite{BST,AETW} 
\begin{eqnarray}\label{LpST}
{\hat{\cal L}}_{p+1}=  
{1\over 2} * \hat{E}_{{a}} \wedge \hat{E}^{{a}} 
- {(p-1)\over 2} (-)^p *1 - \hat{C}_{p+1}
\end{eqnarray}
(explicit expressions for the D$p$-branes and the  M5-brane can be 
found in \cite{c0} and \cite{blnpst}), where  
\begin{equation}\label{hEa}
  \hat{E}^{ {a}} = d\hat{Z}^{ {M}} E_{ {M}}^{~ {a}}(\hat{Z})
= d\xi^m  \partial_m \hat{Z}^{ {M}}(\xi )  E_{ {M}}^{~ {a}}(\hat{Z})\; ,    
\end{equation}  
$\hat{C}_{p+1}\equiv {1\over (p+1)!} d\hat{Z}^{ {M}_{p+1}}\ldots  
d\hat{Z}^{ {M}_1} C_{ {M}_1 \ldots  {M}_{p+1}}(\hat{Z})$ 
are the pull--backs 
$\hat{\phi}^*(E^a)$, $\hat{\phi}^*(C_{p+1})$
of the {\sl bosonic} supervielbein and gauge superforms to $W^{p+1}$, 
and $*$ is the Hodge operator for a ($p$+$1$)--dimensional space  
with (independent) 
worldvolume metric $g_{mn}$ on $W^{p+1}$;    
$* \hat{E}_{{a}} \wedge \hat{E}^{ {a}}= d^{p+1}\xi \sqrt{|g|}g^{mn}  
\hat{E}^{a}_m \hat{E}^b_{n} \eta_{ab}$, 
$\; (-)^p*1$= $d^{p+1}\xi$ $\sqrt{|g|}$.

\medskip

{\bf 3.} {\it  
Equations for the coupled system and fermionic 
fields $\tilde{\theta}$, $\hat{\theta}$}.  
The variational problem associated with the full action (\ref{SSG+p})
requires that the variations of the GM action are extended  
to an integral over superspace $\Sigma^{(D|n)}$,  
\begin{equation}\label{dLD} 
\int_{M^{D}} \delta\tilde{{\cal L}}_D = 
\int_{{\Sigma}^{(D|n)}} d^n\theta (\theta -\tilde{\theta}(x))^n 
\delta\tilde{{\cal L}}_D \; ,
\end{equation}
with the use of the 
Grassmann delta 
function $\delta 
(\theta -\tilde{\theta}(x))= (\theta -\tilde{\theta}(x))^n$,  
and similarly for 
the variations of the superbrane action on $W^{p+1}$, for which 
\begin{eqnarray}\label{dLp} 
& \int_{W^{p+1}} \delta\hat{{\cal L}}_{p+1} = 
\nonumber \\ 
& \int_{{\Sigma}^{(D|n)}} d^D x d^n\theta 
\int_{W^{p+1}}
\delta\hat{{\cal L}}_{p+1}\delta^D (x -\hat{x}(\xi))
(\theta -\hat{\theta}(\xi))^n \, . 
\end{eqnarray}

The variation of (\ref{SSG+p}) with respect to the 
supervielbein coefficient  
$\delta {E}^{ {a}}_{M}(Z)$ produces the equation 
\begin{eqnarray}\label{MDdX=JT} 
& (\theta -\tilde{\theta}(x))^n 
\tilde{M}_{(D-1)~{ {a}}} \wedge d\tilde{Z}^{M}=
d^Dx \times   \nonumber \\  
& \times 
\int_{W^{p+1}} *\hat{E}_a 
\wedge d\hat{Z}^{M} 
(\theta -\hat{\theta}(\xi))^n  \delta^D (x -\hat{x}(\xi)) \; ,   
\end{eqnarray}
where  $\tilde{M}_{(D-1)~{ {a}}}$ is given in  eq. (\ref{tMD-1}).
Factoring out $dx^\mu$,  
eq. (\ref{MDdX=JT}) 
for $M=\mu$  implies 
 \begin{eqnarray}\label{MDdX=JTmu} 
& (\theta -\tilde{\theta}(x))^n 
\tilde{M}_{(D-1)~{ {a}}} 
=  dx^{\wedge (D-1)}_\mu \times \\  
& \times 
\int_{W^{p+1}} *\hat{E}_a 
\wedge d\hat{x}^{\mu} 
(\theta -\hat{\theta}(\xi))^n  \delta^D (x -\hat{x}(\xi))\, .  \nonumber
\end{eqnarray}
The superspace coordinate 
$\theta$ appears in the Grassmann deltas only. 
Thus eq. (\ref{MDdX=JTmu}) contains a set of $n$ $\theta$--independent 
equations corresponding to the $\theta^n$, $(\theta)^{(n-1)}_{\alpha}\equiv
{1\over (n-1)! }
\epsilon_{\alpha\alpha_1 \ldots \alpha_ {(n-1)}}\theta^{\alpha_1}\ldots 
\theta^{\alpha_{(n-1)}}$ 
etc. coefficients in the expansion. The $\theta^n$ coefficient gives  
\begin{eqnarray}\label{Ei=JT} 
&\tilde{M}_{(D-1){ {a}}} = J^{(p)}_{(D-1)a} \equiv \nonumber \\ 
& \equiv (dx)^{\wedge (D-1)}_\mu 
\int_{W^{p+1}} *\hat{E}_a 
\wedge d\hat{x}^{\mu} 
\delta^D (x -\hat{x})\, , 
\end{eqnarray} 
which constitutes 
the `superform' generalization of the {\sl Einstein equation with 
source}. 
The $(\theta)^{(n-1)}_{\alpha}$ equation, after eq. 
(\ref{Ei=JT}) is taken into account, acquires the form 
(we factorize out the common $(D-1)$--form multiplier 
and use 
$\tilde{\theta}^{ {\alpha}}(x) \delta^D (x -\hat{x}(\xi))\equiv 
\tilde{\theta}^{ {\alpha}}(\hat{x}) 
\delta^D (x -\hat{x}(\xi))\, $)
 \begin{eqnarray}\label{MDdTh=JT} 
&  \int_{W^{p+1}} *\hat{E}_a 
\wedge d \hat{x}^\mu \;  
(\hat{\theta}^{ {\alpha}}(\xi )- 
\tilde{\theta}^{ {\alpha}}(\hat{x}(\xi)))\delta^D (x -\hat{x})=0 \, .
\end{eqnarray}
Integrating over $M^D$ with an arbitrary probe function 
$f(x,\tilde{\theta}(x))$ we 
conclude that eq. (\ref{MDdTh=JT}) implies 
\begin{equation}\label{TTT2} 
*\hat{E}_a \wedge d \hat{x}^\mu \;  
( \tilde{\theta}^{ {\alpha}}(\hat{x})-
\hat{\theta}^{ {\alpha}}(\xi))=0 \; 
\end{equation}
and, as the worldvolume is assumed to be a nondegenerate 
({\rm rank}$(\hat{E}^a_m)=(p+1)$) surface, that  
\begin{eqnarray}\label{TTT5} 
\tilde{\theta}^{ {\alpha}}(\hat{x}(\xi))=
\hat{\theta}^{ {\alpha}}(\xi ) \; .   
\end{eqnarray}
Hence, the fermionic coordinate field 
$\tilde{\theta}^{ {\alpha}}(x)$, which plays an auxiliary 
r\^ole in the GM formulation of free supergravity, becomes identified 
on ${\cal W}^{p+1}$ with the worldvolume fermionic coordinate field 
$\hat{\theta}^{ {\alpha}}(\xi )$  in the interacting case, {\it i.e.} 
$\tilde{\phi}\vert_{W^{p+1}}=\hat{\phi}$ 
so that ${\cal W}^{p+1}\; \subset \; {\cal M}^{D}$ and,  
in particular, 
\begin{equation}\label{TTT6}  
\hat{\tilde{E}}^a\equiv
d\tilde{Z}^M(\hat{x}) E_M^a(\hat{x}, 
\tilde{\theta}(\hat{x}))= 
\hat{E}^a\equiv d\hat{Z}^M E_M^a(\hat{x},\hat{\theta})\; . 
\end{equation}  
We note that after replacing $\tilde{M}_{(D-1)a}$ in Eq. 
(\ref{MDdX=JTmu}) by the {\it r.h.s.} of Eq. 
(\ref{Ei=JT}) and using the identification (\ref{TTT5}),  
Eq. (\ref{MDdX=JTmu}) becomes an identity and thus it does not produce other 
independent equations beyond (\ref{Ei=JT}) and (\ref{TTT5}).

The ($p$+$1$)--form gauge field equation (\ref{tG}) 
evidently acquires a source term 
from the super--p--brane Wess--Zumino term $\hat{C}_{p+1}$ (\ref{LpST})
(see \cite{SUGRA} for the bosonic case). 

In our super--$p$--brane---supergravity interacting system, 
$w^{ab}$, $F_{a_1\ldots a_q}$, as well as 
the fermionic supervielbein form ${\hat E}^{\underline{\alpha}}$, 
do not appear in the super-$p$-brane 
action since the Wess--Zumino term $\hat{C}_{p+1}$ is treated as 
the pull--back $\hat{\phi}^*(C_{p+1})$ 
to $W^{p+1}$ 
of an {\sl independent} ($p$+$1$)--form $C_{p+1}$ on $\Sigma^{(D|n)}$.
Hence, eqs. (\ref{tTa}), (\ref{tH}) as well as 
the fermionic superform equation 
(\ref{tRS}) 
will not acquire a source term from $\delta S_{D,p}$.

\medskip 

{\bf 4.} {\it A toy model of the coupled system. 
Local symmetries}. 
Let us consider, for simplicity, the supergravity---superparticle 
coupled system 
(eq. (\ref{SSG+p}) for $p$=$0$) with
\begin{equation}\label{L1}
\hat{{\cal L}}_1 = 
{1 \over 2} e(\tau ) \hat{E}^a 
  \hat{E}^b_{\tau}\eta_{ab}\; , \qquad 
*\hat{{E}}_a= e(\tau)\hat{{E}}_{\tau a}\; . 
\end{equation} 
The superparticle equations of motion 
$(\delta S / \delta \hat{Z}^M(\tau))$ $E_M^a(\hat{Z})$= $0$, 
$(\delta S / \delta \hat{Z}^M(\tau))E_M^{\underline{\alpha}}(\hat{Z})$= $0$, 
$\delta S / \delta e(\tau)= 0$ are 
\begin{eqnarray}\label{pb}
{\cal D}(e(\tau) \hat{E}_{\tau a}) + e(\tau) \hat{E}_{\tau b}
\hat{E}^B T_{B a}{}^b(\hat{Z})=0\; , 
\\ 
\label{pf} 
e(\tau) \hat{E}_{\tau a}
\hat{E}^B T_{B \underline{\alpha}}{}^a(\hat{Z})=0\; ,
\\ 
\label{pe} \hat{{E}}^{a}_{\tau } \hat{E}_{\tau a} =0\; .
\end{eqnarray} 
The gauge field equation (\ref{tG})  
remains `free', as the fermionic equation 
(\ref{tRS}) and the geometric equations (\ref{tTa}), (\ref{tH}) do. 
Thus, the terms denoted by dots in (\ref{DRS}), 
${\cal D} \tilde{M}_{(D-1) a}+ \ldots$=$0$ and 
similar equations vanish. Then the integrability condition 
for eq. (\ref{tRS}) reads ({\it cf.} (\ref{DRS})) 
\begin{equation}\label{tDRS0}
{\cal D} \tilde{\Psi}_{(D-1)\underline{\alpha}} = 
2i \tilde{M}_{(D-1) a}\wedge \tilde{E}^{\underline{\beta}}
\Gamma^a_{\underline{\alpha}\underline{\beta}} =0 \; .  
\end{equation} 
Using now (\ref{Ei=JT}) for $p$=$0$ one finds from  (\ref{tDRS0})
(see (\ref{L1}))
\begin{eqnarray}
\label{curJ}
J^{p=0}_{(D-1) a}\wedge \tilde{E}^{\underline{\beta}}
\Gamma^a_{\underline{\alpha}\underline{\beta}} =0 \; 
\end{eqnarray}
{\it i.e.}, $\,(dx)^{\wedge (D-1)}_\mu
\wedge \tilde{E}^{\underline{\beta}}
\Gamma^a_{\underline{\beta}\underline{\alpha}}
\int_{W^{1}} e \hat{{E}}_{\tau a} 
\, d\hat{x}^{\mu} 
\delta^D (x -\hat{x}) =0$. Due to eqs. 
(\ref{TTT5}), (\ref{TTT6}) this is then equivalent to 
\begin{eqnarray}\label{fEqm2}
\int_{W^{1}} e \hat{{E}}_{\tau a}
\, \hat{{E}}^{\underline{\beta}}
\Gamma^a_{\underline{\beta}\underline{\alpha}} 
\delta^D (x -\hat{x}) = 0\; ,
\end{eqnarray}
which implies 
\begin{equation}\label{fEqm}
\hat{{E}}^{\underline{\beta}}
\Gamma^a_{\underline{\alpha}\underline{\beta}} \hat{E}_{\tau a} =0 \; .
\end{equation} 
Thus eq. (\ref{fEqm}), which follows from   
eq. (\ref{tDRS0}), has the same form as the 
fermionic equation for the superparticle in a superspace supergravity 
background being subject 
to the constraints (\ref{Ta}). 
In the same way the integrability condition for eq. (\ref{Ei=JT})
produces ${\cal D}(e(\tau ) \hat{{E}}_{\tau a})=0$. 

The supergravity part of the coupled system is invariant under 
the local supersymmetry (\ref{lsusy}) due to the identity (\ref{DRS}) 
(now no longer a Noether identity for the coupled system since
it contains $\tilde{M}_{(D-1) a}$ rather than 
$\tilde{M}_{(D-1)a}-J^{(p=0)}_{(D-1)a} $). For the superparticle 
part we have  
\begin{eqnarray}\label{dlsS}
& \delta_{ls} S_{D,0}=  
\int_{W^1} (e(\tau) \hat{E}_{\tau a}\,  \delta_{ls}\hat{E}^a + 
  d\tau {1\over 2} 
\hat{E}_{\tau a}\hat{E}_{\tau}^a \, \delta_{ls} e(\tau)) =
\nonumber \\ 
& = \int_{W^1} (-2i e(\tau) \hat{E}_{\tau a} 
\hat{E}^{\underline{\alpha}}
{\Gamma}^a_{\underline{\alpha}\underline{\beta}} 
\epsilon^{\underline{\beta}}
 +  d\tau {1\over 2} 
\hat{E}_{\tau a} \hat{E}_{\tau}^a\, \delta_{ls} e(\tau))\; , 
\end{eqnarray}
since $\delta_{ls}\hat{E}^a$ is simply given by (\ref{lsusy}) for hatted 
variables  ($\epsilon^{\underline{\alpha}}(\hat{x}, \hat{\theta})$ etc.), 
due to  (\ref{lsusyx}). 
Thus $\delta_{ls} S=0$ if 
\begin{equation}\label{E1/2} 
\hat{\phi}^* (\epsilon^{\underline{\alpha}})\equiv 
\epsilon^{\underline{\alpha}}(\hat{x},\hat{\theta})= \hat{E}_\tau^a
{\Gamma}_a^{\underline{\alpha}\underline{\beta}}{\kappa}_{\underline{\beta}}
(\tau)\; ,
\end{equation}
on ${\cal W}^1$ and $\delta_{ls}e(\tau)=4i \hat{E}_\tau^{\underline{\alpha}}
{\kappa}_{\underline{\alpha}}(\tau)$. 
This shows that the local supersymmetry 
$\delta_{ls}$ (\ref{lsusy}) 
in the supergravity---superbrane coupled system is preserved 
on  ${\cal M}^D$ but not on  
${\cal W}^{1}$ 
where, due to the relation $(\hat{E}_\tau^a \,  
{\Gamma}_a)^2= \hat{E}_\tau^a\hat{E}_{\tau a}=0$,  
it reduces to $(n/2)$ 
 `$\kappa$--like' transformations 
(\ref{E1/2}) \cite{kappa}.

The coupled action evidently possesses 
the local symmetries (\ref{susy}) 
supplemented by the transformations 
$\delta_{\tilde{s}}\hat{Z}^M = b^M (\hat{x})$ 
of the superbrane variables 
\begin{eqnarray}\label{susyW} 
\delta_{\tilde{s}} \hat{x}^\mu= b^\mu (\hat{x}) \, , 
\qquad  
\delta_{\tilde{s}} \hat{\theta}^{\alpha}= 
{\varepsilon}^{\alpha}(\hat{x})\, .
\end{eqnarray}
This invariance is a consequence of the invariance of 
the differential forms $E^A$ etc. under 
superdiffeomorphisms, the pull--backs of which 
give (\ref{susy}) and (\ref{susyW}).

In spite of the local symmetry  (\ref{susy}) for the coupled system, 
the equations cannot be lifted to 
$\Sigma^{(D|n)}$ (see the end of Sec. 2a). Indeed,  
the na\"{\i}ve lifting of eqs. (\ref{tTa}),  (\ref{tRS}) 
produces the superspace constraints (\ref{Ta}) and eq. (\ref{RS}), 
which implies the lifting of (\ref{tDRS0}), $M_{(D-1) a}$=$0$ 
and hence no source. In contrast, eq. (\ref{curJ}) on an arbitrary 
surface ${\cal M}^D$ implies eq. (\ref{fEqm}) with 
$J^{(p=0)}_{(D-1)a}\ne 0$. Thus, the equations of motion for the 
interacting system are formulated in terms of `superfields' 
depending on  $\tilde{\theta}(x)$, {\it i.e.} living on 
the $D$--dimensional surface ${\cal M}^D$.

\medskip

{\bf 5.} {\it Final remarks.} 
The local symmetry 
(\ref{susy}), (\ref{susyW}) of the coupled action 
might suggest that both the bosonic and fermionic 
degrees of freedom of the superbrane $\hat{x}, \hat{\theta}$ as well as 
the supergravity fermion $\tilde{\theta}(x)$ are 
pure gauge. However, the `gauge' $\tilde{\theta}(x)=0$, 
$\hat{\theta}(\xi)=0$ is singular 
(the corresponding superspace transformation is {\sl not} 
a superdiffeomorphism); 
the same applies to setting ${\theta}=0$ in any superfield supergravity.  
Nevertheless, the GM action makes sense when 
 $\tilde{\theta}(x)=0$ is imposed.
In this case eq. (\ref{TTT5}) implies $\hat{\theta}(\xi)=0$ 
and the coupled action 
reduces to that of the {\it bosonic}  
brane plus the component action for supergravity.  
These considerations will apply to  any 
superdiffeomorphism invariant action for the coupled system.

The $D=11$ supergravity 
GM action is known 
\cite{rheo}. It could be directly used 
to the study 
of su\-per\-gravity---M2-brane system. 
An important technical problem for further applications  
of our approach  is the explicit  construction  of 
the group--manifold action for the $D=10$ type IIA and IIB 
su\-per\-gravities  
as well as for the duality invariant version of $D=11$ supergravity 
\cite{bbs}.  
We conclude by noting that the identification 
of the supergravity and superbrane 
fermionic coordinate functions 
(\ref{TTT5}) indicates  
that our approach could also be useful 
in the  Lagrangian description of interacting superbrane systems  
({\it cf.} \cite{BK}).

\medskip

{\it Acknowledgments}. 
The authors are grateful to E. Bergshoeff and M. Cederwall 
for useful conversations 
and comments and especially to D. Sorokin for 
numerous discussions and constructive criticism which led to a much 
clearer presentation. 
This work has been partially supported by the 
DGICYT research grant PB 96-0756, the Ministerio de Educaci\'on 
y Cultura (I.B.), the Generalitat Valenciana and KBN grant 5 P03B 05620 
(J.L.) and the Junta de Castilla y Le\'on (research grant C02/199).

\end{multicols}

\begin{thebibliography}{99}
\bibitem{rheo1} 
This is the content of the rheonomic principle 
\cite{rheo}.
\bibitem{kappa}
The super--$p$--brane actions in the supergravity background being subject to 
the constraints  (\ref{Ta}), (\ref{H}) are invariant under  
local $n/2$ parametric 
fermionic $\kappa$--transformations $\delta_\kappa$.  
In the superparticle case,  
eq. (\ref{L1}),   
the $\kappa$--symmetry transformations are 
$\delta_\kappa \hat{Z}^M E_M^a (\hat{Z})$=$0$,       
$
\delta_\kappa \hat{Z}^M E_M^{\underline{\alpha}} (\hat{Z})$=$ 
\hat{E}_\tau^a{\Gamma}_a^{\underline{\alpha}\underline{\beta}}
$, 
$
\delta_\kappa e$=$4i 
\hat{E}_\tau^{\underline{\beta}} {\kappa}_{\underline{\beta}}\, .
$
In contrast with $\kappa$--symmetry
$\delta_\kappa\hat{Z}\not=0$, in (\ref{lsusy}), (\ref{E1/2}) 
$\delta_{ls}\tilde{Z}=0=\delta_{ls}\hat{Z}$.
\end{thebibliography}
\end{document}